\documentclass[a4paper,fleqn]{cas-dc}

\usepackage[authoryear,longnamesfirst]{natbib}

\def\tsc#1{\csdef{#1}{\textsc{\lowercase{#1}}\xspace}}
\tsc{WGM}
\tsc{QE}

\usepackage{graphicx}
\begin{document}
\let\WriteBookmarks\relax
\def\floatpagepagefraction{1}
\def\textpagefraction{.001}

\shorttitle{ONCship: A Large-Scale Underwater Acoustic Corpus for Ship Recognition}

\shortauthors{}  

\title[mode=title]{ONCship: A Large-Scale Underwater Acoustic Corpus for Ship Recognition}



%

\author[1,2]{Zhe Cao}
\fnmark[1]
\ead{zcao123@mail.nwpu.edu.cn}

\author[1,2]{Jichao Zhang}
\fnmark[1]
\ead{zhangjichao@mail.nwpu.edu.cn}

\author[1,2]{Xiao-Lei Zhang}
\cormark[1]
\ead{xiaolei.zhang@nwpu.edu.cn}

\fntext[1]{These authors contributed equally to this work and share first authorship.}

\affiliation[1]{
    organization={School of Marine Science and Technology, Northwestern Polytechnical University},
    city={Xi'an},
    postcode={710072},
    country={China}
}

\affiliation[2]{
    organization={Research and Development Institute of Northwestern Polytechnical University in Shenzhen},
    city={Shenzhen},
    postcode={518063},
    country={China}
}

\cortext[1]{Corresponding author}



\begin{abstract}
Recent advances in deep learning have significantly improved underwater acoustic vessel recognition. However, existing publicly available datasets are typically limited in scale, hindering large-scale representation learning and downstream task evaluation. To address this limitation, we introduce ONCship, a large-scale underwater acoustic dataset comprising 4521 hours of curated recordings from Ocean Networks Canada (ONC) hydrophones and Automatic Identification System (AIS) data. We evaluate the proposed dataset across three tasks: self-supervised representation learning, vessel-type classification, and MMSI-Level Vessel Classification. Experiments on DeepShip and ShipsEar demonstrate the effectiveness of ONCship for acoustic representation learning, with joint pre-training on ONCship and AudioSet achieving a classification accuracy of 62.45\% on DeepShip through linear probing. Furthermore, incorporating ONCship as auxiliary training data improves downstream vessel-type classification performance. Beyond conventional vessel-type classification, we further investigate MMSI-Level Vessel Classification, treating each vessel as a distinct identity. Experimental results show that models fine-tuned with different loss functions achieve Top-1 accuracies exceeding 53\% and Top-5 accuracies approaching 69\%, demonstrating the feasibility of distinguishing individual vessels through MMSI-Level Vessel Classification. Overall, ONCship provides a valuable resource for underwater acoustic representation learning and vessel recognition, while its rich vessel-level metadata enables further exploration of MMSI-Level Vessel Classification.
\end{abstract}


\begin{highlights}
\item ONCship contains 4521h of underwater recordings from six ONC stations.
\item Domain-specific pre-training improves DeepShip NMI and retrieval AUC.
\item Pre-training benefits vary across metrics and evaluation datasets.
\end{highlights}


\maketitle

\section{Introduction}
\label{sec:introduction}

Underwater acoustic target recognition uses hydrophone recordings to identify vessel types and other acoustic sources. It supports maritime surveillance, vessel traffic management, fisheries enforcement, and ocean noise monitoring \cite{aslam2024review}. Passive acoustic monitoring does not require illumination and can complement radar and optical sensing, but sound propagation and weather-dependent ambient noise affect the recorded signals \cite{hummel2024survey}. Vessel-type classification describes broad categories, whereas individual-vessel identification can link repeated observations of a known vessel, a finer-grained task analogous to speaker identification in speech processing. Identity recognition should therefore be evaluated across temporally separated encounters rather than neighboring clips from the same source recording. Both task levels require training data that cover this acoustic variation.

Limited annotated data constrain underwater representation learning and evaluation. Widely used benchmarks such as DeepShip \cite{irfan2021deepship} and ShipsEar \cite{santos2016shipsear} together provide only tens of hours, whereas general-audio corpora such as AudioSet contain thousands of hours \cite{gemmeke2017audio}. For supervised vessel-type classification, limited independent labeled recordings also restrict the diversity available during training \cite{huang2026fewshot}. Supplementary recordings can expand this coverage, but their value depends on whether they support the benchmark's recognition task. Larger underwater resources are emerging: UniqueShip contains 2460h of ship-radiated audio from 4218 vessels, or 3437h including background recordings, and separates vessel identities across benchmark partitions \cite{hashemi2026uniqueship}. Its analysis shows that identity overlap can inflate estimates of generalization in vessel-type classification. The Maritime Mobile Service Identity (MMSI) therefore serves both as a grouping key for identity-aware type-classification evaluation and as a label for individual-vessel identification. Expanding labeled training data and defining identity-level tasks both require attention to these metadata and evaluation boundaries.

Self-supervised learning offers another use of large audio archives: learning representations without task-specific labels during pre-training. Underwater applications include masked-token learning \cite{feng2024masking} and contrastive learning on separate unlabeled ONC recordings \cite{hummel2026contrastive}. Their transfer depends on the training distribution: underwater recordings contain low-frequency ship-noise components shaped by propagation and ambient noise \cite{hummel2024survey}, while general-audio corpora cover a broader range of sound events. This difference motivates comparing general-audio, underwater, and joint pre-training. Together with the labeled-data and identity requirements above, it defines three complementary questions: how pre-training sources affect transfer, how additional labeled recordings affect vessel-type classification, and whether acoustic information distinguishes known vessels across temporally separated encounters.

\begin{figure}
\centering
\includegraphics[width=\columnwidth]{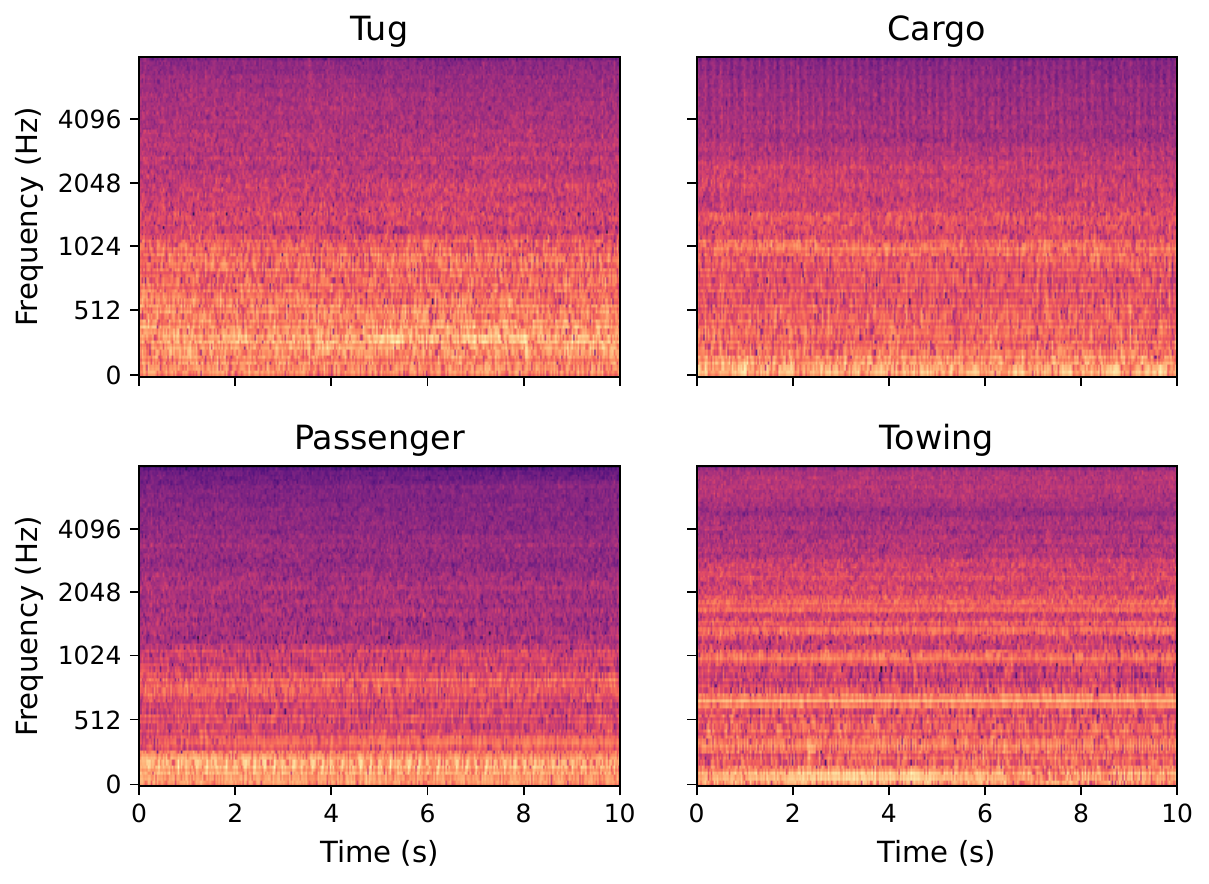}
\caption{Example log-mel spectrograms for the four most frequent vessel types in ONCship: Tug, Cargo, Passenger, and Towing. Each panel shows one recording.}
\label{fig:spectrograms}
\end{figure}

To address these three questions, we construct ONCship from the long-term hydrophone recordings and vessel-traffic observations provided by ONC \cite{heesemann2014ocean}. Automatic Identification System (AIS) messages provide vessel identifiers, types, and positions for association with acoustic recording intervals. Starting from 20000h of raw recordings, trajectory processing, spatial association, and temporal selection yield a 4521h corpus. It includes AIS-associated single-ship and multi-ship intervals, background intervals, and recordings for which AIS data were not downloaded. For self-supervised pre-training, we exclude recordings from the FAE and LSBBL stations used by DeepShip, leaving 3905h of ONCship data. Figure~\ref{fig:spectrograms} presents log-mel spectrogram examples for the four most frequent vessel types in ONCship.

Using ONCship, we evaluate the three questions under distinct protocols. First, we examine how pre-training sources affect representation quality and classification transfer, using the same Efficient Audio Transformer (EAT) architecture and objective for AudioSet-only, ONCship-only, and joint pre-training \cite{chen2024eat}. Second, we assess how supplementary labeled recordings affect DeepShip vessel-type classification, comparing models trained on the benchmark alone with those receiving additional ONCship data from the same or different hydrophones, with supplementary MMSIs entirely disjoint from DeepShip. Third, we test whether acoustic information supports identification of known vessels across temporally separated encounters, using MMSI labels and keeping all clips from the same source recording within one data partition.

The contributions of this study are threefold:
\begin{itemize}
\item We construct ONCship, a 4521h underwater acoustic corpus curating single-ship, multi-ship, and ambient recordings with rich AIS vessel-type and MMSI metadata, providing a large-scale resource for both label-free pre-training and downstream supervised tasks.
\item We systematically evaluate ONCship across self-supervised representation learning, vessel-type classification, and MMSI-Level Vessel Classification. The experiments cover general-audio, domain-specific, and joint pre-training, matched and mismatched hydrophone deployments, and fine-grained discrimination of individual vessels across different voyages and temporal encounters.
\end{itemize}

Section~\ref{sec:related} reviews related work, Section~\ref{sec:construction} describes ONCship construction, and Section~\ref{sec:experiments} reports the pre-training, vessel-type classification, and MMSI identification experiments. Section~\ref{sec:conclusion} summarizes the findings and limitations.

\section{Related Work}
\label{sec:related}

\subsection{Underwater Acoustic Datasets}

Underwater acoustic datasets have evolved from relatively small benchmarks designed for vessel-type classification to larger corpora with richer vessel annotations, more diverse acoustic scenes, and long-term recordings. Existing datasets differ substantially in recording duration, vessel taxonomy, background annotations, acquisition environments, and data construction strategies \cite{aslam2024review}. Representative resources include ShipsEar, DeepShip, QiandaoEar22, Oceanship, IARA, and UniqueShip, while long-term observatories such as ONC provide the raw hydrophone and AIS data from which several large-scale datasets have been constructed.

\textbf{ONC:} ONC operates long-term underwater observatories that provide open-access hydrophone recordings together with AIS information \cite{heesemann2014ocean}. These recordings cover extended periods and contain diverse vessel activities, environmental noise, and complex multi-source acoustic scenes. Several widely used ship-radiated noise datasets, including DeepShip, VTUAD, Oceanship, and UniqueShip, have been constructed from subsets of the ONC archive.

\textbf{ShipsEar:} ShipsEar is a widely used underwater vessel classification dataset collected near the port of Vigo, Spain \cite{santos2016shipsear}. It contains approximately 2.8h of recordings from eleven vessel types. The detailed vessel labels can also be grouped into broader categories, making ShipsEar a common benchmark for evaluating feature extraction and supervised vessel classification methods.

\textbf{DeepShip:} DeepShip is a ship-radiated noise dataset constructed from ONC recordings acquired between 2016 and 2018 \cite{irfan2021deepship}. It contains approximately 47.1h of recordings from vessels passing near the hydrophones and defines four vessel categories: cargo ships, passenger ships, tankers, and tugboats. DeepShip has been widely adopted as a benchmark for underwater acoustic target recognition and vessel classification.

\textbf{QiandaoEar22:} QiandaoEar22 was collected in Qiandao Lake, China, and contains approximately 31.4h of ship and background recordings \cite{du2024qiandaoear22}. The dataset contains 20 target labels and includes both single-target and multi-target acoustic scenes. Its multi-target recordings support recognition tasks in which multiple acoustic sources may be simultaneously present.

\textbf{Oceanship:} Oceanship is a ship-radiated noise dataset curated from the ONC repository using recordings collected between July 2020 and February 2021 \cite{li2024oceanship}. It contains approximately 121.2h of audio and extends the vessel taxonomy beyond the four broad categories commonly used in earlier ONC-derived datasets, providing fifteen vessel-type categories. The dataset combines hydrophone recordings with AIS information to support supervised vessel-type classification.

\textbf{IARA:} IARA contains approximately 129.6h of underwater acoustic recordings collected in the Santos Basin, Brazil \cite{silva2025iara}. It provides annotations for 34 detailed vessel types that are further organized into eight general vessel groups. The dataset also includes background recordings, providing a relatively broad range of vessel categories and underwater acoustic conditions.

\textbf{UniqueShip:} UniqueShip is a large-scale dataset constructed from long-term ONC recordings \cite{hashemi2026uniqueship}. It contains approximately 3437h of audio, including background recordings, and includes recordings associated with 4218 distinct vessels. Its benchmark defines eleven vessel types together with a background class.

\begin{table*}[t]
\centering
\caption{Selected underwater acoustic datasets. Durations are approximate totals in hours, including background where available.}
\label{tab:datasets}
\begin{tabular*}{\textwidth}{@{\extracolsep{\fill}}p{0.32\textwidth}p{0.07\textwidth}p{0.12\textwidth}p{0.17\textwidth}p{0.16\textwidth}@{}}
\hline
Dataset & Year & Duration (h) & Label scope & Source \\
\hline
ShipsEar \cite{santos2016shipsear} & 2016 & 2.8 & 11 vessel types & Vigo, Spain \\
DeepShip \cite{irfan2021deepship} & 2021 & 47.1 & 4 vessel types & ONC, Canada \\
QiandaoEar22 \cite{du2024qiandaoear22} & 2024 & 31.4 & 20 target labels & Qiandao Lake, China \\
Oceanship \cite{li2024oceanship} & 2024 & 121.2 & 15 vessel types & ONC, Canada \\
IARA \cite{silva2025iara} & 2025 & 129.6 & 34 vessel types  & Santos Basin, Brazil \\
UniqueShip \cite{hashemi2026uniqueship} & 2026 & 3437.0 & 11 vessel types & ONC, Canada \\
ONCship (Ours) & 2026 & 4521.0 & 22 vessel types  & ONC, Canada \\
\hline
\end{tabular*}
\end{table*}

\subsection{Underwater Acoustic Target Recognition}

Underwater acoustic target recognition has progressed from handcrafted signal descriptors to learned representations. Traditional systems combine spectral features, including LOFAR line spectra, DEMON envelope spectra, and Mel-frequency cepstral coefficients, with conventional classifiers \cite{aslam2024review}. These descriptors encode properties of ship-radiated noise, but their effectiveness depends on feature selection and recording conditions. Convolutional networks learn patterns in time-frequency inputs \cite{xie2024diversity}, while Transformer-based models capture relationships across the input sequence \cite{wu2025estmst}. This progression shifts more of the representation design into the model, increasing the importance of training data that cover the acoustic variation encountered at inference.

Beyond conventional vessel-type classification, several studies have explored finer-grained vessel recognition from underwater acoustic signals. Karakos et al. formulated individual-ship detection as an open-class detection problem analogous to speaker identification \cite{karakos2018individual}. More recently, Du and Hong studied specific-target identification using the QiandaoEar22 dataset \cite{du2024qiandaoear22}. These studies demonstrate the potential of underwater acoustic signals for recognition beyond coarse vessel-type classification.

Changes to model architecture alone do not remove the constraint of limited independent labeled recordings. Regularization and data augmentation seek to improve robustness \cite{xie2024diversity,aslam2024review}, while few-shot learning targets recognition with fewer labeled examples \cite{huang2026fewshot}. Nevertheless, high-capacity supervised models remain at risk of overfitting when the available data are small or insufficiently diverse \cite{xie2024diversity}. Expanding a network's capacity cannot substitute for expanding the range of acoustic conditions represented during learning. This bottleneck motivates learning representations from the much larger unlabeled archives, with the limited labeled data used to adapt or evaluate those representations on specific recognition tasks.

\subsection{Self-Supervised Audio Pre-Training and Underwater Adaptation}

Self-supervised audio pre-training separates representation learning from task-specific annotation by constructing learning targets from the audio itself. Reconstruction-based methods recover masked spectrogram content \cite{huang2022masked}, whereas token-prediction methods use clustered or learned acoustic units as targets \cite{liu2023dinosr,chen2022beats}. Teacher-student methods instead learn by matching representations of related or masked inputs \cite{baevski2023efficient,chen2024eat}. These objectives impose different constraints on the representation while allowing large unlabeled corpora to be used during pre-training. They reduce dependence on labels at that stage, but the resulting features still require downstream evaluation; successful general-audio pre-training does not establish their suitability for every underwater recognition setting.

Underwater adaptation must account for a training distribution shaped by ship-radiated noise, propagation, and ambient sound rather than speech or general sound events \cite{hummel2024survey}. Domain-specific self-supervised learning has already been investigated: \citet{feng2024masking} apply hierarchical masked-token learning to underwater target recognition, and \citet{hummel2026contrastive} learn contrastive representations from separate unlabeled ONC recordings and evaluate frozen features on downstream ship and marine-mammal classification tasks. These studies establish the feasibility of learning underwater representations without task-specific labels during pre-training. These studies motivate comparing general-audio, underwater-audio, and joint pre-training at thousand-hour scale under a common learning framework.

Addressing this question requires a common model and objective so that the comparison does not also change the recognition architecture. We use the EAT, which combines an efficient self-supervised pre-training framework with prediction of teacher representations at both frame and utterance levels \cite{chen2024eat}. Its general-audio formulation provides a basis for comparing AudioSet, ONCship, and their combination using the same encoder architecture and pre-training objective. This focuses the study on the choice and combination of training data. Downstream evaluation then examines whether any observed benefits extend across benchmarks, metrics, and adaptation protocols, rather than presuming that domain-specific pre-training is uniformly advantageous.

\section{Dataset Construction}
\label{sec:construction}

\begin{figure*}
\centering
\includegraphics[width=\textwidth]{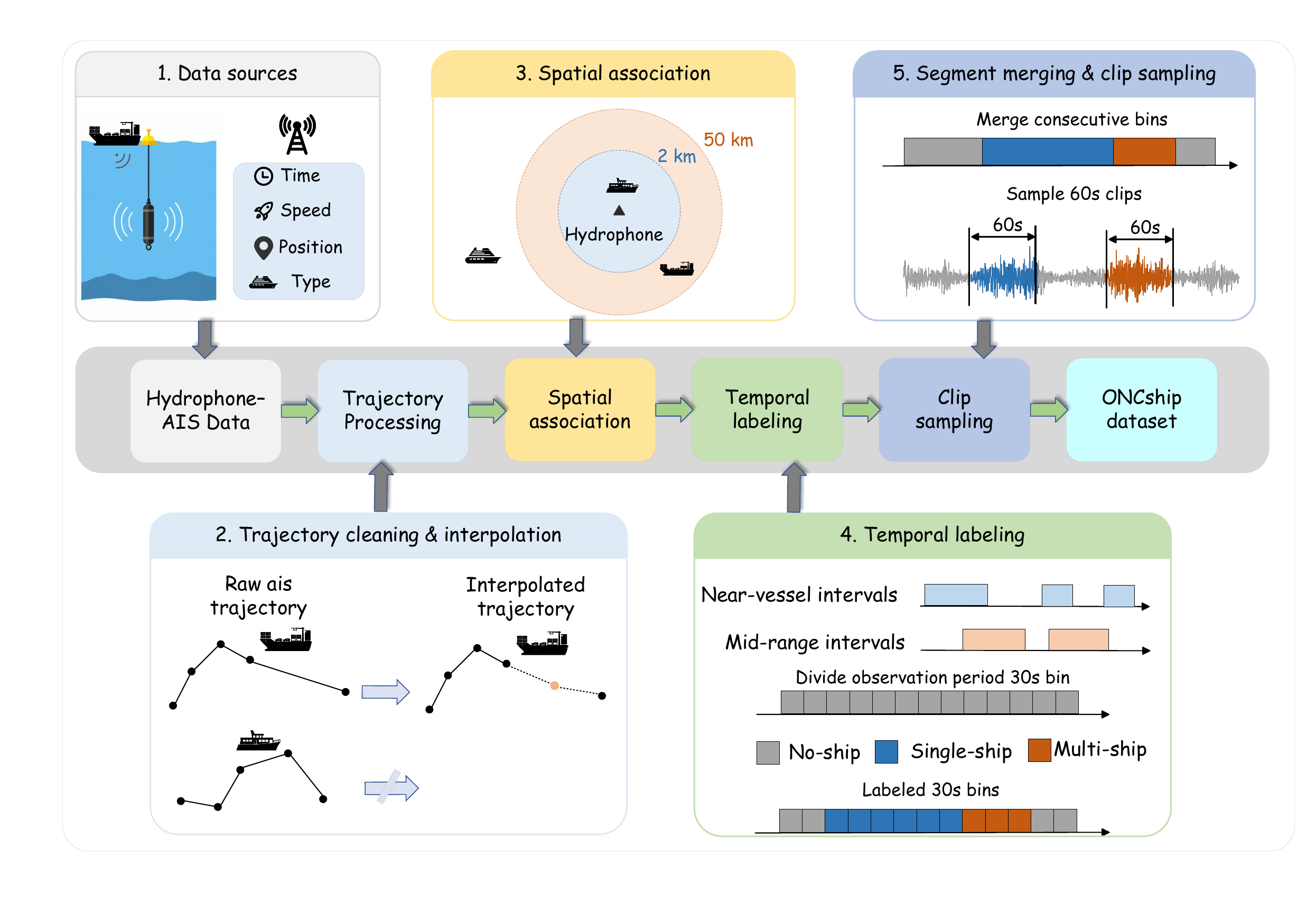}
\caption{ONCship construction pipeline. The five steps collect hydrophone and AIS data, clean and interpolate vessel trajectories, associate trajectories with hydrophones, label 30s time bins, and merge segments before sampling 60 s clips. Sampled clips are subsequently subdivided into 10s clips.}
\label{fig:pipeline}
\end{figure*}

ONCship is constructed by curating an archive of 20000h of raw hydrophone recordings into a 4521h underwater acoustic corpus. Figure~\ref{fig:pipeline} illustrates five steps in this process: hydrophone and AIS data collection, trajectory cleaning and interpolation, spatial association, temporal labeling, and segment merging with clip sampling. The workflow combines acoustic recordings with vessel-traffic observations to identify recording intervals, associate vessel metadata where available, and organize the selected audio into training-ready subsets. Section~\ref{sec:data-source} describes the recording stations and AIS sources, Section~\ref{sec:filtering} details the trajectory-based association and audio selection procedure, and Section~\ref{sec:distribution} reports the corpus composition and vessel-type statistics.

\subsection{Data Source}
\label{sec:data-source}

Recordings are obtained from ONC, a cabled ocean observatory operated by the University of Victoria that maintains instrument platforms in the coastal and offshore waters of British Columbia, Canada, spanning the Strait of Georgia, the waters around Vancouver Island, and the northeast Pacific Ocean \cite{onc_data}. The observatory streams continuous data from underwater sensors, including broadband hydrophones, and pairs these recordings with vessel traffic information received by shore-based AIS stations.

\begin{figure*}
  \centering
  \includegraphics[width=0.32\textwidth]{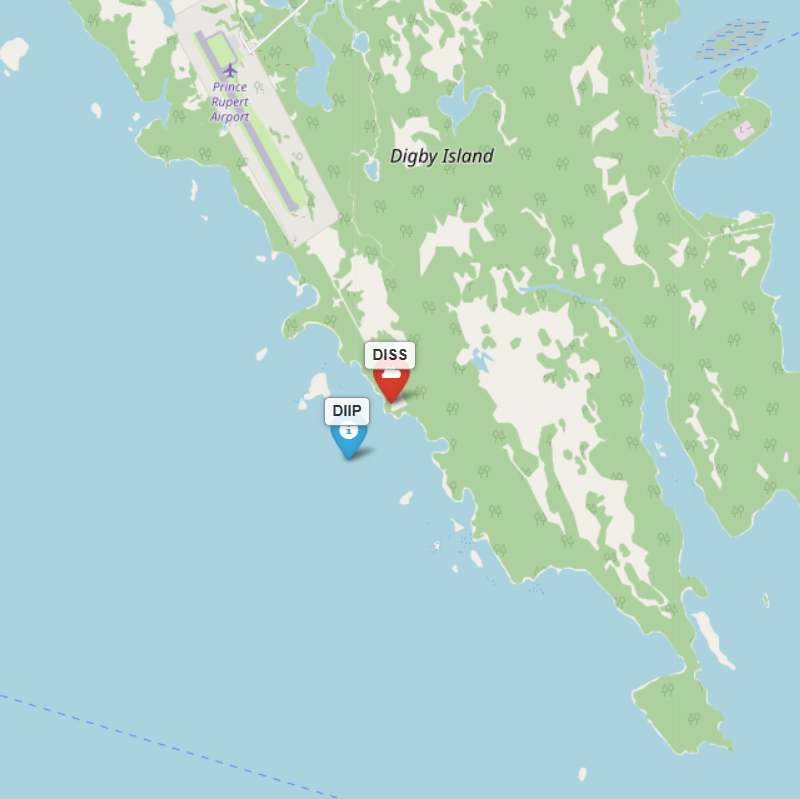}
  \hfill
  \includegraphics[width=0.32\textwidth]{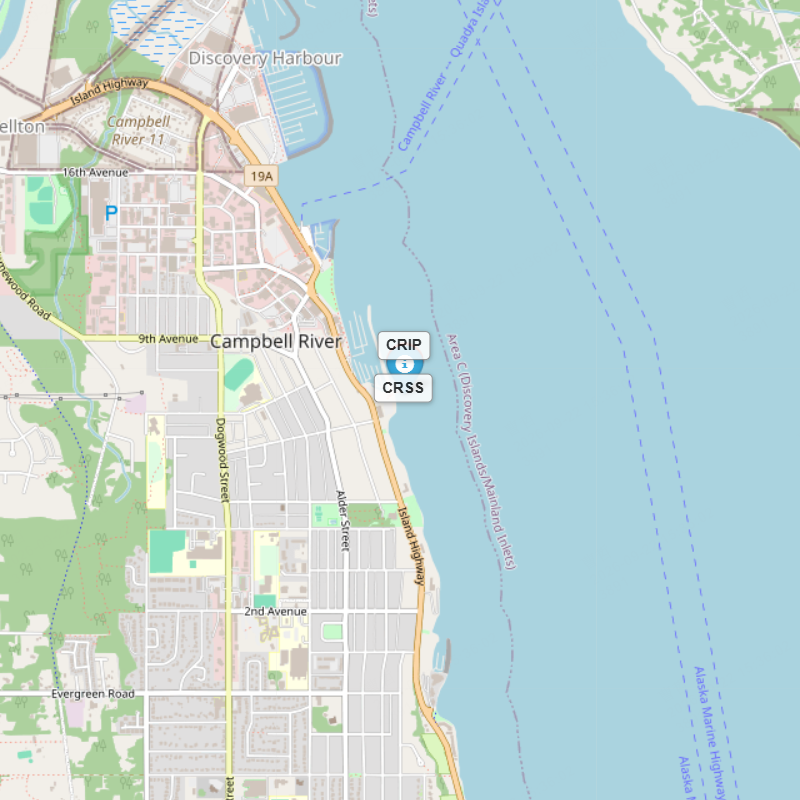}
  \hfill
  \includegraphics[width=0.32\textwidth]{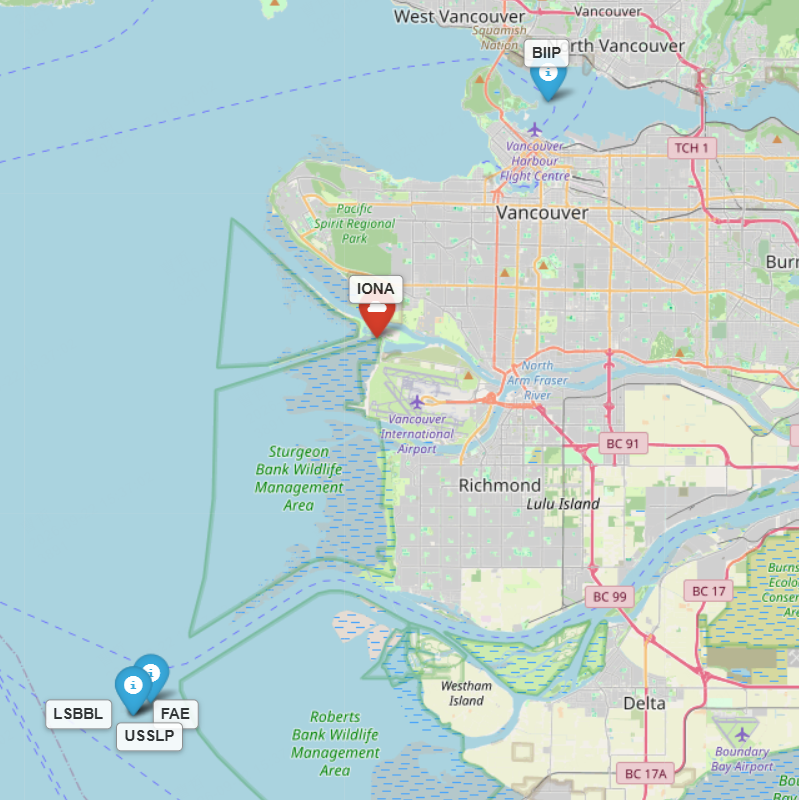}
  \caption{Locations of the ONC hydrophone stations and associated AIS receivers. Left: DIIP and DISS. Center: CRIP and CRSS. Right: the Vancouver-area stations and IONA. Base map: OpenStreetMap contributors, available under the Open Database License (\url{https://www.openstreetmap.org/copyright}).}
  \label{fig:stations}
\end{figure*}

The dataset draws on recordings from six hydrophone stations, identified by the codes DIIP, FAE, LSBBL, BIIP, CRIP, and USSLP, located between $49.08^\circ\,\mathrm{N}$ and $54.26^\circ\,\mathrm{N}$; Figure~\ref{fig:stations} shows their locations together with the AIS receivers. Each station hosts one or more icListen hydrophones deployed over successive periods, and the complete deployment history is listed in Table~\ref{tab:stations}. In total, the recordings span March 2016 to May 2026, with individual deployments lasting from several months to a few years.

All recordings are resampled to a common rate of 32 kHz to standardize the waveform sampling grid across deployments. This rate gives a nominal Nyquist frequency of 16 kHz, encompassing the low-frequency tonal components commonly used in ship-noise analysis \cite{hummel2024survey}. For higher-rate inputs, resampling also reduces the number of samples processed per clip and the associated waveform storage and processing requirements.

Vessel identity and type are obtained from the shore-based AIS receivers (IONA, CRSS and DISS) listed in Table~\ref{tab:stations}. The AIS messages provide the MMSI, vessel type, and position of each transmitting vessel, which we use to associate recordings with specific ships. We selected the USSLP recordings for self-supervised pre-training, which does not require vessel labels, and therefore did not download the corresponding AIS records. These recordings are retained without AIS-derived vessel associations and reported as a separate subset in Section~\ref{sec:distribution}.

\begin{table*}
\centering
\caption{Complete hydrophone deployment history with the associated AIS receivers. Coordinates are decimal degrees; positive latitude denotes north and negative longitude west. A dash denotes that AIS records were not downloaded.}
\label{tab:stations}
\small
\setlength{\tabcolsep}{3pt}
\begin{tabular*}{\textwidth}{@{\extracolsep{\fill}}llllll}
\hline
Station & Lat., Lon. ($^\circ$) & Device ID & Recording period & AIS & AIS Lat., Lon. ($^\circ$) \\
\hline
\multirow{9}{*}{DIIP} & \multirow{9}{*}{$54.259, -130.431$} & ICLISTENHF1352 & 2016.03.30--2017.04.06 & \multirow{9}{*}{DISS} & \multirow{9}{*}{$54.263, -130.425$} \\
 & & ICLISTENHF1352 & 2017.04.07--2018.10.25 & & \\
 & & ICLISTENHF1354 & 2018.10.26--2019.08.29 & & \\
 & & ICLISTENHF1352 & 2019.08.29--2021.06.02 & & \\
 & & ICLISTENHF1293 & 2021.06.02--2021.10.12 & & \\
 & & ICLISTENHF1293 & 2021.10.12--2022.07.16 & & \\
 & & ICLISTENHF1351 & 2022.07.16--2023.09.16 & & \\
 & & ICLISTENHF6406 & 2023.09.16--2025.09.12 & & \\
 & & ICLISTENHF1252 & 2025.09.12--2026.05.13 & & \\
\hline
\multirow{2}{*}{FAE} & \multirow{2}{*}{$49.081, -123.339$} & ICLISTENAF2523 & 2017.06.24--2017.11.03 & \multirow{2}{*}{IONA} & \multirow{2}{*}{$49.216, -123.205$} \\
 & & ICLISTENAF2523 & 2017.11.04--2018.10.04 & & \\
\hline
LSBBL & $49.081, -123.339$ & ICLISTENAF2523 & 2016.05.02--2017.06.24 & IONA & $49.216, -123.205$ \\
\hline
\multirow{5}{*}{BIIP} & \multirow{5}{*}{$49.301, -123.111$} & ICLISTENHF1951 & 2020.10.22--2022.06.17 & \multirow{5}{*}{IONA} & \multirow{5}{*}{$49.216, -123.205$} \\
 & & ICLISTENHF1354 & 2022.10.04--2023.07.25 & & \\
 & & ICLISTENHF1354 & 2023.07.25--2024.02.06 & & \\
 & & ICLISTENHF1354 & 2024.02.06--2025.01.24 & & \\
 & & ICLISTENHF1354 & 2025.01.25--2026.05.13 & & \\
\hline
\multirow{9}{*}{CRIP} & \multirow{9}{*}{$50.021, -125.235$} & ICLISTENHF1350 & 2016.03.18--2016.07.28 & \multirow{9}{*}{CRSS} & \multirow{9}{*}{$50.021, -125.235$} \\
 & & ICLISTENHF1350 & 2016.07.29--2017.01.24 & & \\
 & & ICLISTENHF1350 & 2017.01.26--2018.01.30 & & \\
 & & ICLISTENHF1333 & 2018.02.14--2018.07.17 & & \\
 & & ICLISTENHF1333 & 2018.11.15--2020.07.14 & & \\
 & & ICLISTENAF2523 & 2020.07.15--2022.01.28 & & \\
 & & ICLISTENHF1350 & 2022.01.28--2024.03.18 & & \\
 & & ICLISTENHF1349 & 2024.03.19--2025.03.24 & & \\
 & & ICLISTENHF6017 & 2025.08.01--2026.05.13 & & \\
\hline
USSLP & $49.085, -123.330$ & ICLISTENAF2556 & 2018.10.04--2019.02.28 & - & - \\
\hline
\end{tabular*}
\end{table*}

\subsection{Data Selection}
\label{sec:filtering}

The 20000h archive contains vessel activity and ambient underwater sound. For stations with downloaded AIS records, we associate recording intervals with vessel trajectories and select single-ship, multi-ship, and no-ship segments using the rules below. USSLP recordings are retained separately because their AIS records were not downloaded. This subsection describes the AIS-based procedure that produces the audio clip process plan.

For each hydrophone, dynamic AIS records within a 50km radius are collected and grouped by MMSI into single-vessel trajectories. Trajectories of stationary vessels are discarded. A trajectory is treated as stationary when it contains fewer than two points, its mean speed-over-ground is below 1 knot, at least 80\% of its points fall below 1 knot, or its end-to-end displacement is less than 0.3km.

For the remaining trajectories, intermediate points are inserted at 30s intervals when the gap between consecutive AIS observations lies between 30s and 30 min. Each scalar attribute $q$ (latitude, longitude, or speed-over-ground) is interpolated separately as $q(t)=q_i+\alpha(q_{i+1}-q_i)$, where $q_i$ and $q_{i+1}$ are its observed values at times $t_i$ and $t_{i+1}$, and $\alpha=(t-t_i)/(t_{i+1}-t_i)$. Interpolation times are $t=t_i+k\tau$, with $\tau=30$ s and positive integers $k$ satisfying $t_i<t<t_{i+1}$; the endpoints remain the original AIS observations. Surface distances between the vessel positions and the hydrophone coordinates are then computed using the Haversine formula, with latitude and longitude converted to radians.

Two kinds of vessel-presence intervals are then extracted. Close-range intervals are continuous segments within 2km of the hydrophone; among them, only segments whose duration does not exceed 1.5h, whose mean speed-over-ground is at least 3 knots, and whose proportion of points below 3 knots does not exceed 40\% are retained and marked as vessel-present. Mid-range intervals are continuous segments whose distance to the hydrophone lies between 2km and 50km; these are marked as mid-range vessel present without further filtering. The 2km association radius is chosen to favor recordings in which vessel sound is more distinguishable from ambient noise, since a shorter propagation path generally reduces transmission loss. It is a signal-to-noise consideration rather than a measured acoustic detection boundary: received levels also depend on vessel source strength and local propagation conditions \cite{silva2025iara}.

The observation period is divided into 30s time bins. A bin covered by a retained close-range interval is added to the near-vessel set, and a bin covered by a mid-range interval is marked as mid-range. Bins with no near or mid-range vessel are labeled no-ship; bins with no near vessel but a mid-range vessel are excluded; bins with exactly one near vessel are labeled single-ship and record its MMSI, mean speed, and type; bins with more than one near vessel are labeled multi-ship and record the vessel count. Excluding mid-range-only bins reduces the risk of weaker, more distant shipping sounds entering the background negative samples even when no close-range vessel is present. The no-ship label is defined by these AIS-based rules, not by verified absence of all shipping noise. Single-ship and multi-ship refer to the number of retained near-vessel trajectories, not to all vessels within 50km.

Consecutive bins with identical status, and, for single-ship bins, identical MMSI, whose time gap does not exceed 30s, are merged into segments. Segments from different devices and days of the same hydrophone are then combined: no-ship and multi-ship segments are merged directly, while single-ship segments are divided into those with and without valid vessel-type information. Two segments are merged when the end time of one equals the start time of the next and all metadata except time are identical. The total duration of each category is then accumulated.

Fixed-duration audio clips are sampled randomly from the merged timeline for each category. Within a segment of duration $L$, a clip of duration $D=60$ s must lie entirely inside the segment, so its start offset is sampled from $[0,L-D]$ when $L\geq D$. Accepted clip start times within the segment are separated by at least $2D=120$s. For single-ship segments, at most 1500 clips are sampled per MMSI. The output is a per-category clip list with total durations, which forms the audio clip download.

\subsection{Dataset Distribution}
\label{sec:distribution}

We further subdivided the extracted 60s segments into 10s clips. Table~\ref{tab:duration} presents the final duration of each category in the dataset. Specifically, single-ship recordings with vessel-type metadata account for 3106.8 hours, multi-ship recordings for 929.1 hours, single-ship recordings without vessel-type metadata for 38.4 hours, and no-ship recordings for 7.4 hours. The 439.3 hours of recordings from USSLP are reported separately, as the corresponding AIS data were not downloaded. Together, these categories comprise a total of 4521 hours of underwater acoustic recordings.

Table~\ref{tab:types} presents all 24 vessel-type labels from the original metadata. After excluding the Reserved and Unknown labels from the class count and supervised classification, 22 vessel types remain, as reported in Table~\ref{tab:datasets}. Among these, Tug, Cargo, Passenger, and Towing account for the majority of clips with vessel-type metadata. However, the number of clips varies by several orders of magnitude across vessel types, resulting in a pronounced long-tailed distribution.

\begin{figure}
\centering
\includegraphics[width=\columnwidth]{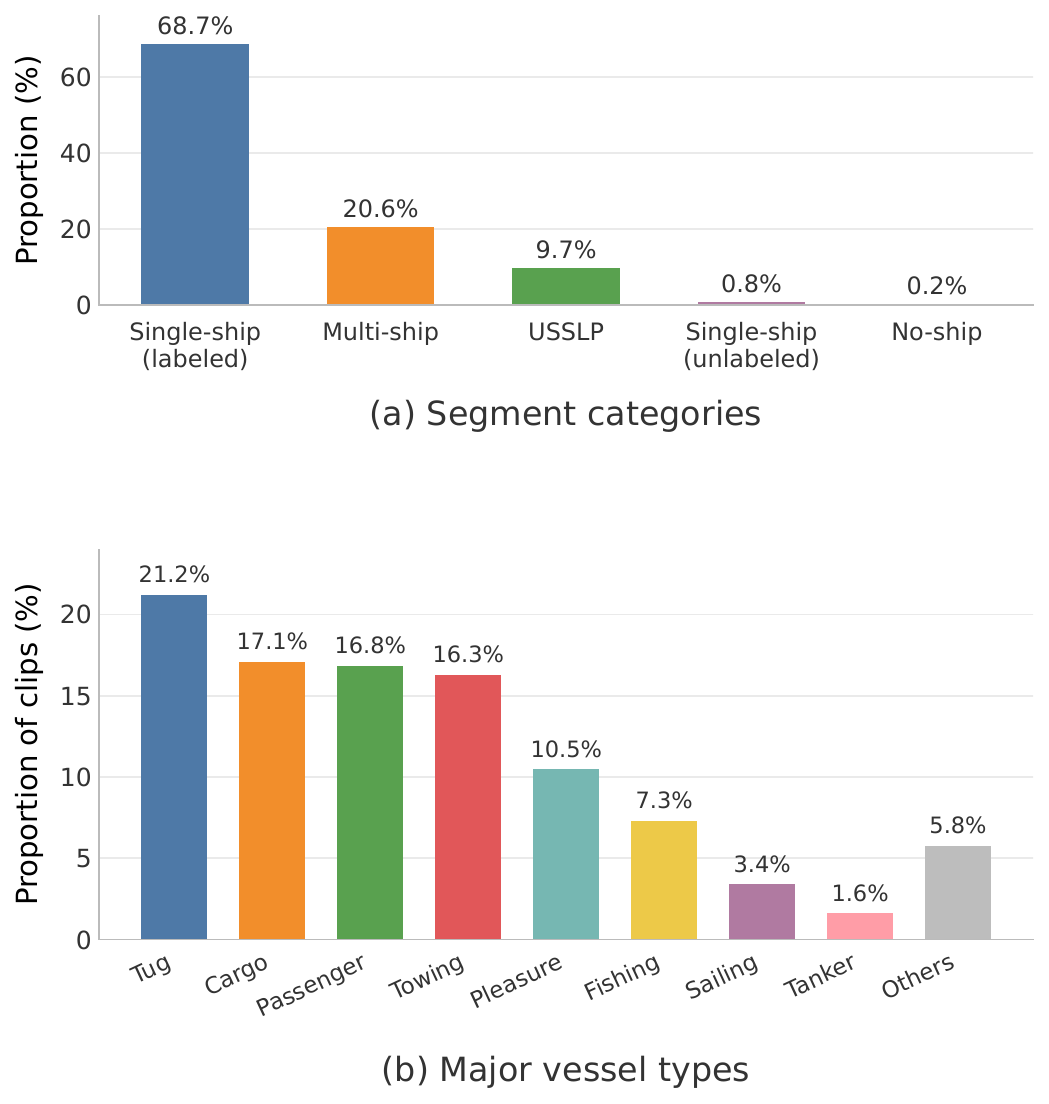}
\caption{ONCship distribution. (a) Duration by recording category. (b) Major vessel-type labels among single-ship recordings with type metadata; Others aggregates all remaining labels, including Reserved and Unknown.}
\label{fig:distribution}
\end{figure}

\begin{table}
\centering
\caption{Total duration of each segment category after filtering and clipping.}
\label{tab:duration}
\begin{tabular*}{\columnwidth}{@{\extracolsep{\fill}}lc}
\hline
Category & Duration (h) \\
\hline
Single-ship, labeled & 3106.8 \\
Single-ship, unlabeled & 38.4 \\
Multi-ship & 929.1 \\
No-ship & 7.4 \\
USSLP (AIS not downloaded) & 439.3 \\
\hline
\end{tabular*}
\end{table}

\begin{table}
\centering
\caption{Number of 10 s clips per raw vessel-type metadata label. SAR: search and rescue; HSC: high-speed craft; WIG: wing-in-ground craft; RR18: noncombatant vessels covered by Radio Regulations Resolution No.~18 (AIS ship-type code 59) \cite{guo2023ais}.}
\label{tab:types}
\begin{tabular*}{\columnwidth}{@{\extracolsep{\fill}}lc}
\hline
Metadata label & Clips \\
\hline
Tug & 237238 \\
Cargo & 191287 \\
Passenger & 188183 \\
Towing & 182301 \\
Pleasure & 117036 \\
Fishing & 81616 \\
Sailing & 38115 \\
Tanker & 17978 \\
Pilot & 14888 \\
SAR & 12942 \\
Other & 10928 \\
Port tender & 5307 \\
Local & 5003 \\
Military & 3834 \\
Reserved & 3232 \\
Anti-pollution & 2828 \\
Dredging & 1666 \\
Law enforcement & 1187 \\
Diving & 1074 \\
HSC & 1012 \\
WIG & 515 \\
Unknown & 126 \\
Medical & 120 \\
RR18 & 18 \\
\hline
\end{tabular*}
\end{table}

\section{Experiments}
\label{sec:experiments}
In this section, we first introduce the evaluation metrics and then conduct experiments from three perspectives to demonstrate the effectiveness of our dataset. First, we investigate whether large-scale underwater vessel recordings can improve the feature representation capability of pretrained models. Second, we examine whether the collected data can improve downstream vessel-type classification performance. Finally, we evaluate the feasibility of the fine-grained MMSI-Level Vessel Classification task, which aims to distinguish individual vessels rather than only coarse vessel types.

\subsection{Evaluation Metrics}
\label{sec:metrics}

We evaluate classification performance using accuracy (ACC), top-5 accuracy (Top-5 ACC), weighted F1-score (Weighted-F1), macro F1-score (Macro-F1), and macro recall (Macro-Recall). ACC measures the overall proportion of correctly classified samples, while Top-5 ACC considers a prediction correct if the ground-truth class is among the five highest-scoring classes. Top-5 ACC is only reported for the MMSI-Level Vessel Classification task, where the number of classes exceeds five.

For class $c$, let $TP_c$, $FP_c$, and $FN_c$ denote the numbers of true positives, false positives, and false negatives. The class-wise recall and F1-score are
\begin{equation*}
\mathrm{Recall}_c=\frac{TP_c}{TP_c+FN_c}, \qquad
\mathrm{F1}_c=\frac{2TP_c}{2TP_c+FP_c+FN_c}.
\end{equation*}
Macro-Recall and Macro-F1 are obtained by averaging the corresponding class-wise scores equally over all classes, while Weighted-F1 weights each class according to its number of samples. These metrics complement ACC by providing a more balanced evaluation under class imbalance.

For frozen representation evaluation, we report normalized mutual information (NMI) and retrieval AUC. NMI measures the agreement between K-means cluster assignments of the extracted features and the ground-truth labels. Retrieval AUC is computed from pairwise cosine similarities between frozen representations. Each sample is treated as a query, with samples from the same class regarded as positives and samples from different classes as negatives. A ROC-AUC score is computed for each valid query and then averaged:
\begin{equation*}
\mathrm{AUC}=\frac{1}{|\mathcal{Q}|}\sum_{i\in\mathcal{Q}}\mathrm{AUC}_i,
\end{equation*}
where $\mathcal{Q}$ denotes queries containing both positive and negative candidates. This AUC therefore evaluates the discriminability of the learned representation rather than classifier probabilities.

\subsection{Models}
\label{sec:baselines}

To evaluate the effectiveness and generalizability of ONCship, we conduct a series of experiments using established and competitive audio learning architectures across different training paradigms and downstream tasks. Rather than proposing a new model architecture, our objective is to investigate how large-scale underwater acoustic data can benefit audio representation learning and downstream underwater sound recognition. Specifically, we select EAT and BEATs as representative transformer-based architectures, and include a previously reported convolutional baseline for additional comparison.

\textbf{Efficient Audio Transformer (EAT)}  \cite{chen2024eat} is a self-supervised audio representation learning framework based on teacher-student bootstrapping and the Utterance-Frame Objective (UFO). We adopt EAT to investigate the effectiveness of ONCship under three different pre-training configurations, as described in Section~\ref{sec:pretraining}.

\textbf{BEATs} \cite{chen2022beats} is a self-supervised audio pre-training framework that iteratively learns acoustic tokenizers and audio representations through masked discrete token prediction. We use BEATs to examine whether incorporating ONCship recordings improves downstream fine-tuning performance.

\textbf{ResNet} \cite{he2016deep} is a convolutional neural network
architecture that employs residual connections to facilitate
the training of deep networks. We include ResNet-based models
as convolutional baselines to complement the transformer-based
architectures. Specifically, ResNet18 is used for comparison
with previously reported contrastive representation learning
results, whereas ResNet50 is employed as a supervised baseline
for evaluating the effectiveness of additional ONCship
recordings in vessel-type classification.

\subsection{Effectiveness of the Dataset for Self-Supervised Pre-Training}
\label{sec:pretraining}

To evaluate the effectiveness of pre-training on different datasets, we compare three EAT models trained from scratch: EAT-as, pre-trained exclusively on AudioSet; EAT-onc, pre-trained exclusively on ONCship; and EAT-as\&onc, pre-trained on the combination of both datasets. We assess the quality of the learned audio representations on downstream classification tasks by freezing the pre-trained encoders and training lightweight classifiers on their extracted features. This evaluation allows us to directly examine the contribution of different pre-training datasets without the influence of end-to-end fine-tuning. Additionally, we include the DeepShip and ShipsEar classification accuracy reported for LARS, a contrastive learning method proposed by \citet{hummel2026contrastive}, as an external reference.

All three configurations use the EAT architecture \cite{chen2024eat}, with a 12 layer transformer and a hidden dimension of 768. The AudioSet portion contains 5000h \cite{gemmeke2017audio}; the ONCship portion contains 3905h after excluding all recordings from FAE and LSBBL, the stations represented in DeepShip. This exclusion is by station, not by recording date, and applies to the pre-training corpus. The joint configuration contains 8905h. All pre-training audio is divided into 10s clips. Each model is randomly initialized and does not load an initial pre-trained checkpoint.

We adopt the Utterance-Frame Objective (UFO) proposed by \citet{chen2024eat} for self-supervised pre-training, using an exponential-moving-average (EMA) teacher-student framework. The training objective combines frame-level and utterance-level mean-squared errors (MSE). The spectrogram patch-masking ratio is set to 0.8, and no additional data augmentation is applied during pre-training.

Training uses Adam with a cosine learning-rate schedule, a peak learning rate of $5\times10^{-4}$, 75000 warm-up steps, a batch size of 128, bf16 precision, and 30 epochs. Waveforms are resampled to 16 kHz and converted into 128-bin log-mel spectrograms using a 25 ms window and a 10 ms frame shift.

To evaluate the quality of the learned representations, we freeze the pre-trained encoder and apply temporal average pooling to obtain a clip-level feature vector. A multinomial logistic regression classifier is trained on the extracted features, with the regularization parameter $C$ set to 1.0. For DeepShip, we follow the chronological split proposed by \citet{hummel2026contrastive}, using recordings collected before December 2017 for training and those collected from December 2017 onward for testing. We further ensure that the training and test sets are disjoint at the vessel level, with no vessel appearing in both partitions. For ShipsEar, we adopt an 8:2 split based on vessel identity, similarly ensuring vessel-level independence. Audio recordings are subsequently divided into 30s segments with a 15s overlap within each partition to prevent data leakage across the training and test sets.

As shown in Table~\ref{tab:linear}, pre-training on ONCship alone achieves higher NMI and AUC than AudioSet on DeepShip, suggesting that domain-specific underwater acoustic data can improve the discriminative quality of learned representations. Combining AudioSet and ONCship further improves performance, achieving the highest AUC of 0.6540 and ACC of 62.45\% on DeepShip. On ShipsEar, EAT-as achieves the best NMI and AUC, while EAT-as\&onc obtains the highest ACC of 76.18\%, outperforming EAT-as by 5.78 percentage points. These results indicate that ONCship provides complementary acoustic information to general-purpose audio datasets, and that joint pre-training can improve downstream classification performance, although the benefits vary across datasets and evaluation metrics.

Table~\ref{tab:contrastive} further compares the EAT models with the contrastive learning baselines reported by \citet{hummel2026contrastive}. All three EAT models outperform the ResNet18 baselines in terms of both accuracy and weighted F1-score on DeepShip and ShipsEar. In particular, EAT-as\&onc achieves the best overall performance, with an accuracy of 62.45\% and a weighted F1-score of 62.49\% on DeepShip, exceeding ResNet18 (LARS) by 4.67 and 6.49 percentage points, respectively. The improvements are more pronounced on ShipsEar, where EAT-as\&onc achieves 76.18\% accuracy and 77.25\% weighted F1-score, surpassing ResNet18 (LARS) by 32.24 and 35.25 percentage points, respectively. These results further demonstrate the effectiveness of EAT-based self-supervised pre-training and suggest that incorporating ONCship alongside general-purpose audio data can yield competitive representations for downstream underwater acoustic classification.

\begin{table*}[t]
\centering
\caption{Comparison of EAT models pre-trained from scratch on different datasets.
NMI, AUC, and ACC denote normalized mutual information, feature-retrieval AUC,
and classification accuracy, respectively. The best result in each column is highlighted in bold.}
\label{tab:linear}
\begin{tabular*}{\textwidth}{@{\extracolsep{\fill}}lcccccc}
\hline
\multirow{2}{*}{Model}
& \multicolumn{3}{c}{DeepShip}
& \multicolumn{3}{c}{ShipsEar} \\
\cline{2-7}
& NMI & AUC & ACC (\%) & NMI & AUC & ACC (\%) \\
\hline
EAT-as
& 0.1877 & 0.6297 & 61.75
& \textbf{0.4280} & \textbf{0.7284} & 70.40 \\

EAT-onc
& \textbf{0.2021} & 0.6437 & 61.35
& 0.3727 & 0.7127 & 62.36 \\

EAT-as\&onc
& 0.1993 & \textbf{0.6540} & \textbf{62.45}
& 0.3882 & 0.7264 & \textbf{76.18} \\
\hline
\end{tabular*}
\end{table*}

\begin{table*}[t]
\centering
\caption{Comparison of downstream classification performance
between EAT models and contrastive learning baselines.
ACC and Weighted F1 denote classification accuracy and
weighted F1-score, respectively.
The best reported result in each column is highlighted in bold.}
\label{tab:contrastive}
\begin{tabular*}{\textwidth}{@{\extracolsep{\fill}}lcccc}
\hline
\multirow{2}{*}{Model}
& \multicolumn{2}{c}{DeepShip}
& \multicolumn{2}{c}{ShipsEar} \\
\cline{2-5}
& ACC (\%) & Weighted F1 (\%) & ACC (\%) & Weighted F1 (\%) \\
\hline
ResNet18 (Adam)
& 53.94 & 49.00 & 21.40 & 8.00 \\

ResNet18 (LARS)
& 57.78 & 56.00 & 43.94 & 42.00 \\

EAT-as
& 61.75 & 61.93 & 70.40 & 70.85 \\

EAT-onc
& 61.35 & 61.13 & 62.36 & 61.85 \\

EAT-as\&onc
& \textbf{62.45} & \textbf{62.49} & \textbf{76.18} & \textbf{77.25} \\
\hline
\end{tabular*}
\end{table*}

\subsection{Effectiveness of the Dataset for Fine-Tuning}
\label{sec:finetuning}

To further investigate the utility of ONCship in downstream classification tasks, we evaluate its contribution to supervised fine-tuning using three representative architectures: ResNet50, EAT, and BEATs. We consider three training configurations: DeepShip only, DeepShip augmented with ONCship recordings from different hydrophone stations, and DeepShip augmented with ONCship recordings from the same hydrophone stations. These configurations allow us to investigate the contribution of additional underwater acoustic data and the influence of hydrophone sources on downstream classification performance.

All models are fine-tuned end-to-end with a linear classification head for 10 epochs using the Adam optimizer, a batch size of 16, a learning rate of $2\times10^{-5}$, and a weight decay of $10^{-6}$. The final-epoch checkpoint is used for evaluation. DeepShip is partitioned by vessel identity to ensure that recordings from the same vessel do not appear in both training and test sets. Audio recordings are subsequently segmented into 30s clips with a 15s overlap within each partition. For the augmented configurations, additional ONCship recordings are incorporated into the training set according to their hydrophone sources, while the DeepShip test set remains unchanged. All experiments are repeated three times, and the mean and standard deviation are reported. 

Table~\ref{tab:finetune} presents the fine-tuning results under the three training configurations. BEATs consistently achieves the highest performance across all evaluation metrics, reaching an accuracy of 69.41\% when augmented with ONCship recordings from the same hydrophone stations. For EAT, incorporating ONCship recordings from the same stations improves accuracy from 65.64\% to 68.16\%, while recordings from different stations yield only a marginal improvement to 65.95\%. Similarly, BEATs benefits from additional ONCship data, with accuracy increasing from 68.11\% to 68.85\% and 69.41\% for different and same hydrophone sources, respectively. In contrast, ResNet50 exhibits a decline in performance when additional ONCship data are introduced. These results suggest that ONCship can enhance the fine-tuning performance of pre-trained audio transformers, with the benefits depending on model architecture and the acoustic characteristics of the additional training data.

\begin{table*}[t]
\centering
\caption{Fine-tuning performance on DeepShip under different training data configurations.
ACC, F1, and Recall are reported as percentages (mean $\pm$ standard deviation over three runs).
The highest mean within each training configuration is highlighted in bold.}
\label{tab:finetune}
\scriptsize
\setlength{\tabcolsep}{2pt}
\begin{tabular*}{\textwidth}{@{\extracolsep{\fill}}lccccccccc}
\hline
\multirow{2}{*}{Model}
& \multicolumn{3}{c}{DeepShip}
& \multicolumn{3}{c}{ONCship (Different Hydrophones)}
& \multicolumn{3}{c}{ONCship (Same Hydrophones)} \\
\cline{2-10}
& ACC (\%) & macro-F1 (\%) & macro-Recall (\%)
& ACC (\%) & macro-F1 (\%) & macro-Recall (\%)
& ACC (\%) & macro-F1 (\%) & macro-Recall (\%) \\
\hline
ResNet50
& $39.45 \pm 7.94$
& $30.53 \pm 7.03$
& $36.78 \pm 5.76$
& $37.63 \pm 6.35$
& $30.00 \pm 8.02$
& $36.31 \pm 7.82$
& $36.38 \pm 6.86$
& $28.89 \pm 9.31$
& $34.93 \pm 8.30$ \\

EAT-as
& $65.64 \pm 5.96$
& $64.08 \pm 4.82$
& $65.38 \pm 4.24$
& $65.95 \pm 5.16$
& $64.39 \pm 3.52$
& $64.92 \pm 3.37$
& $68.16 \pm 3.75$
& $65.74 \pm 1.75$
& $66.74 \pm 1.70$ \\

BEATs
& $\mathbf{68.11 \pm 3.95}$
& $\mathbf{65.65 \pm 2.64}$
& $\mathbf{65.99 \pm 2.68}$
& $\mathbf{68.85 \pm 4.43}$
& $\mathbf{66.30 \pm 3.25}$
& $\mathbf{66.86 \pm 2.78}$
& $\mathbf{69.41 \pm 2.22}$
& $\mathbf{67.03 \pm 0.25}$
& $\mathbf{67.83 \pm 1.27}$ \\
\hline
\end{tabular*}
\end{table*}

\subsection{Effectiveness of the Dataset for MMSI-Level Vessel Classification}
\label{sec:mmsi}

\begin{figure*}
\centering
\includegraphics[width=\textwidth]{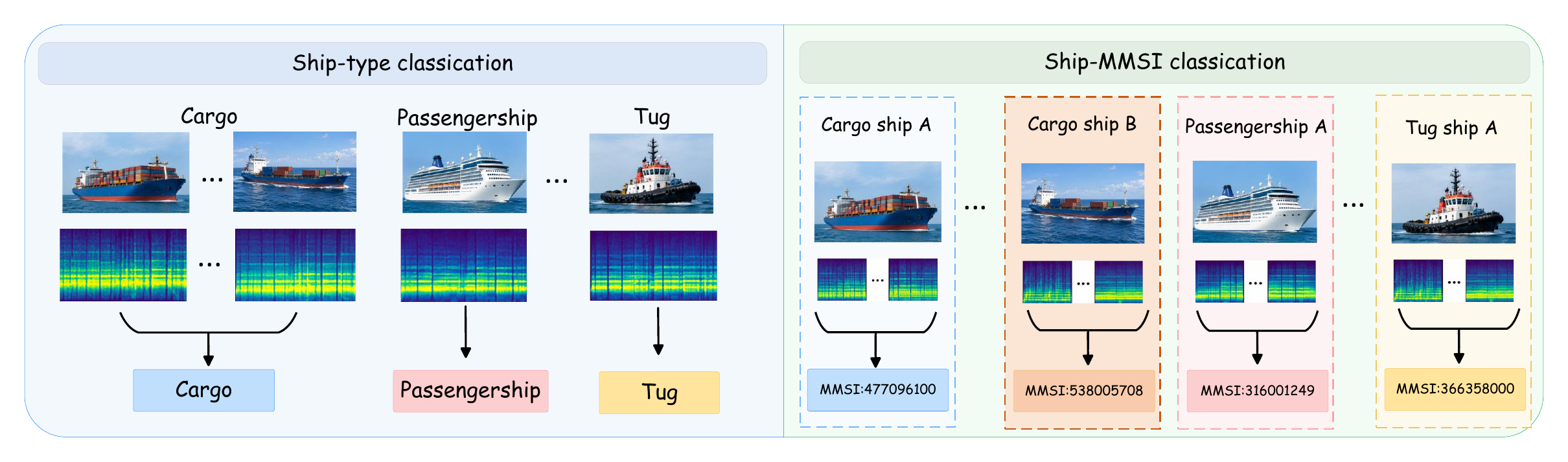}
\caption{Comparison between conventional vessel-type classification and MMSI-Level Vessel Classification. Conventional vessel-type classification assigns underwater acoustic recordings from different vessels to coarse-grained vessel categories (left), whereas MMSI-Level Vessel Classification aims to identify individual vessels based on their MMSI labels (right). Recordings from the same vessel share a common MMSI label, while different vessels of the same type are treated as distinct identities.}
\label{fig:mmsi}
\end{figure*}

To further explore the potential of fine-grained acoustic modeling and fully exploit the rich maritime metadata provided in ONCship, we further investigate MMSI-level vessel classification, where individual vessels are treated as distinct recognition targets.
Unlike conventional underwater acoustic classification benchmarks that categorize targets into broad vessel types, this task aims to identify specific vessel instances by capturing their unique acoustic signatures associated with MMSI numbers. 
As illustrated in Figure~\ref{fig:mmsi} (left), conventional vessel-type classification groups different vessels into coarse-grained categories. In contrast, MMSI-Level Vessel Classification, shown in Figure~\ref{fig:mmsi} (right), aims to assign acoustic segments from the same vessel across different voyages and timestamps to a unified identity class.

For this evaluation, we curate a benchmark subset comprising 50 distinct MMSIs with a cumulative duration of 100 hours. 
For each MMSI, we ensure that pairwise 1 minute source segments are gathered from different timestamps and distinct voyages separated by at least 6 hours. 
Each 1 minute source segment is subsequently segmented into 10-second audio clips. 
To strictly prevent data leakage, all 10-second clips derived from the same 1 minute source segment are constrained to reside exclusively within either the training, validation, or test set. 
We benchmark this task using the EAT-as model, systematically assessing the capability of deep acoustic representations to capture vessel-specific acoustic characteristics in the presence of complex acoustic propagation and ambient oceanic noise.

For EAT-as, a linear classification head was appended to the pre-trained backbone for downstream adaptation. The model was trained for 10 epochs using the Adam optimizer, with a batch size of 64, an initial learning rate of $2 \times 10^{-5}$, and a weight decay of $1 \times 10^{-6}$. We explored two loss functions, cross-entropy and ArcFace\cite{deng2019arcface}, with the latter configured using an angular margin of $m=0.2$ and a scale factor of $s=16.0$. For each configuration, the checkpoint achieving the highest validation accuracy was selected for final evaluation.

\begin{table*}
\centering
\caption{ MMSI-Level Vessel Classification results.
EAT(CE) denotes the model trained with cross-entropy loss,
while EAT(ArcFace) employs ArcFace loss. The best result in each column is highlighted in bold.}
\label{tab:mmsi}
\begin{tabular*}{\textwidth}{@{\extracolsep{\fill}}lccc}
\hline
Model & Top-1 ACC & Top-5 ACC & Macro-F1 \\
\hline
EAT(CE) & 0.5312 & \textbf{0.6957} & 0.5307 \\
EAT(ArcFace) & \textbf{0.5320} & 0.6892 & \textbf{0.5333} \\
\hline
\end{tabular*}
\end{table*}

Table~\ref{tab:mmsi} presents the MMSI-Level Vessel Classification results obtained by fine-tuning with cross-entropy and ArcFace losses. Both loss functions yield comparable performance, with Top-1 accuracies exceeding 0.53 and Macro-F1 scores of approximately 0.53. Specifically, ArcFace slightly outperforms CE in terms of Top-1 accuracy and Macro-F1, achieving 0.5320 and 0.5333, respectively, whereas CE achieves marginally higher Top-5 accuracy, reaching 0.6957. The Top-5 accuracies of both methods approach 0.70, indicating that the correct vessel identity is ranked among the top five candidates in most cases. These findings suggest that underwater acoustic signals contain discriminative features associated with individual vessels, demonstrating the feasibility of MMSI-Level Vessel Classification while highlighting the potential for further improvements in fine-grained vessel classification.

\section{Conclusion}
\label{sec:conclusion}

To address the scarcity of large-scale labeled data for underwater acoustic vessel recognition, this paper constructs ONCship, a large-scale underwater acoustic dataset based on hydrophone recordings from ONC and AIS data. Approximately 4521 hours of valid recordings were curated from nearly 20000 hours of raw acoustic data, covering diverse vessel types and marine environments. Experiments on DeepShip and ShipsEar demonstrate that ONCship can improve acoustic representation learning, while using it as auxiliary training data further benefits downstream vessel-type classification. We also investigate MMSI-Level Vessel Classification, where individual vessels are treated as distinct identities. The results show Top-1 accuracies above 53\% and Top-5 accuracies close to 70\%, demonstrating the feasibility of distinguishing individual vessels from underwater acoustic signals. Overall, ONCship provides a scalable resource for large-scale underwater acoustic representation learning, vessel-type classification, and fine-grained vessel identification. Future work will focus on improving the robustness and generalization of acoustic representations across different hydrophones, marine environments, and noise conditions.


\bibliographystyle{cas-model2-names}

\bibliography{cas-refs}



\end{document}